\documentclass[sigconf,screen,nonacm,review=false,timestamp=true]{acmart}

\usepackage{subcaption}
\usepackage[table]{xcolor}
\usepackage{multirow}
\usepackage[most]{tcolorbox}

\newtcolorbox{promptbox}[1]{
  colback=violet!3,
  colframe=black,
  boxrule=0.5pt,
  arc=2pt,
  left=6pt,
  right=6pt,
  top=6pt,
  bottom=6pt,
  fontupper=\ttfamily\small,
  title=\textbf{#1},
  coltitle=black,
  attach title to upper,
  boxed title style={
    colback=orange!10,
    boxrule=0pt,
    borderline south={0.5pt}{0pt}{black},
  }
}

\AtBeginDocument{%
  }

\setcopyright{acmlicensed}
\copyrightyear{2026}
\acmYear{2026}
\acmDOI{XXXXXXX.XXXXXXX}
\acmConference[FSE '26]{34th ACM Symposium on the Foundations of Software Engineering}
  {June 5--9, 2026}
  {Montreal, Canada}
\acmBooktitle{Companion Proceedings of the 34th ACM Symposium on the Foundations of Software Engineering (FSE '26), June 5--9, 2026, Montreal, Canada}

\acmISBN{978-1-4503-XXXX-X/2018/06}

\begin{document}

\title{Beyond Summaries: Structure-Aware Labeling of Code Changes with Large Language Models}

\author{Bar Weiss}
\authornote{Work done while interning at IBM Research.}
\affiliation{%
  \institution{Viterbi Faculty of Electrical and Computer Engineering}
  \city{Technion, Haifa}
  \country{Israel}}
\email{barweiss@campus.technion.ac.il}

\author{Antonio Abu-Nassar}
\affiliation{%
  \institution{IBM Research}
  \city{Haifa}
  \country{Israel}}
\email{Antonio.Abu.Nassar@ibm.com}

\author{Adi Sosnovich}
\affiliation{%
  \institution{IBM Research}
  \city{Haifa}
  \country{Israel}}
\email{adisos@il.ibm.com}

\author{Karen Yorav}
\affiliation{%
  \institution{IBM Research}
  \city{Haifa}
  \country{Israel}}
\email{yorav@il.ibm.com}

\renewcommand{\shortauthors}{Bar Weiss, Antonio Abu-Nassar, Adi Sosnovich, Karen Yorav}

\begin{abstract}
Code review is a critical practice in software engineering, yet the growing scale and frequency of code patches in modern projects, together with the widespread adoption of AI code assistants, make manual review increasingly challenging. Identifying the types of changes within a patch, such as renames, moves, or logic modifications, can substantially improve review efficiency by enabling prioritization, filtering, and automation. However, existing LLM-based approaches to code review have largely focused on summarization and comment generation, leaving structured code reviews underexplored. In this paper, we present a systematic study of using large language models (LLMs) for taxonomy-based labeling of code changes in a code patch. We introduce a two-stage pipeline that assigns labels to diff hunks and then refines them to capture structural relationships and semantic attributes, such as rename propagation and type changes. Our approach employs few-shot prompting to produce language-agnostic and customizable labels, without the engineering overhead of traditional static-analysis pipelines. We evaluate four LLMs across multiple context configurations on a manually curated benchmark of natural and synthetic patches. Our best configuration achieves up to $84\%$ recall and $81\%$ precision, with high accuracy in extracting relational and attribute metadata. These results suggest that LLM-based labeling can effectively complement static analysis by enabling flexible, multilingual, and automation-friendly code review workflows.
\end{abstract}





\maketitle
\begin{table}
\vspace{0.7cm}
  \centering
  \setlength{\tabcolsep}{10pt}      
  \renewcommand{\arraystretch}{1.3} 

  \begin{tabular}{| l | c | c |}
    \hline
    \rowcolor{black}
    \textcolor{white}{} &
    \textcolor{white}{\textbf{Static Analysis}} &
    \textcolor{white}{\textbf{LLM}} \\
     \hline
    Language Agnostic      & \textcolor{red}{No}                & \textcolor{green!50!black}{Yes} \\
    \hline
    Development Effort     & \textcolor{red}{Very High}         & \textcolor{green!50!black}{Low}  \\
    \hline
    Maintenance            & Medium                             & \textcolor{green!50!black}{Low}  \\
    \hline
    Provable Guarantees    & \textcolor{green!50!black}{Yes}    & \textcolor{red}{No}              \\
    \hline
    Label Customization    & Medium                             & \textcolor{green!50!black}{High} \\
    \hline
    Accuracy               & \textcolor{green!50!black}{High}   & ?                                \\
    \hline
  \end{tabular}
    \caption{Static analysis and LLM properties comparison.}
    \label{tab:static-vs-llm}
    \vspace{-0.4cm}
\end{table}
\section{Introduction}\label{sec:intro}

Code review is an essential activity in software engineering: having human peers inspect changes helps catch defects early, ensures consistent standards, and fosters shared knowledge among team members \cite{sadowski2018modern}. Yet, as codebases grow and code patches become more frequent, review throughput and quality increasingly become bottlenecks. We use the term \emph{code patch} to denote a set of code changes. In practice, such patches are often obtained from pull requests (PRs), but our formulation is independent of the surrounding review workflow. Recent advances in Large Language Models (LLMs) have transformed code-related tasks, including code completion, bug detection, and summarization \cite{brown2020language, kojima2022large, wei2021finetuned}. In the context of code review, LLMs are increasingly used to summarize patches, generate review comments, and explain changes \cite{sun2024llms, codedog2024, palos2024feedback}. While these approaches aim to reduce cognitive load and improve productivity, they primarily produce free-text outputs, which are difficult to integrate into automated workflows or analytics pipelines.

One promising direction is to support both human understanding and automation by labeling diff hunks according to a well-defined taxonomy of change types (e.g., rename, move, interface change, style change). Such structured output enables downstream post-processing, supports tailored workflows, and integrates naturally with dashboards and analytics systems. Traditionally, change-type identification relies on static analysis of program structure, offering strong accuracy and formal guarantees within supported domains. However, these methods are often language-specific, costly to extend, and expensive to maintain. By contrast, an LLM-based approach promises greater flexibility: it can be largely language-agnostic (assuming the model has sufficient exposure to the language), easier to adapt to new taxonomies, and faster to deploy. This flexibility comes at the cost of reduced formal guarantees and potentially less predictable accuracy, motivating a careful empirical study of its capabilities and limitations.

This paper presents a study of LLMs for structured labeling of diff hunks in patches. Specifically, we investigate whether LLMs can:
(1) assign taxonomy-based labels to individual diff hunks; (2) capture structure-aware relationships across hunks (e.g., renaming a function argument and all of its usages); and (3) extract semantic attributes (e.g., old/new identifier names for renaming changes, old/new types for retyping changes).
To evaluate our approach, we assembled a manually curated benchmark of PRs and assessed four LLMs across multiple retrieval-context strategies. Our best configuration achieves 84\% recall and 81\% precision. Overall, our method complements classical static-analysis techniques by trading formal guarantees for faster adoption, multilingual support, and easier customization, while retaining structured outputs that enable practical downstream automation. We conclude by discussing trade-offs, limitations, and opportunities for extending this framework to semantic and intent-level change classification.

\footnote{Additional experimental details and prompt templates are available in the  an anonymized appendix at \url{https://figshare.com/s/a254f611ba26b4da18a2}.}

\section{Related Work}
Traditional approaches to identifying fine-grained code changes rely on static analysis of program structure. These techniques typically operate on Abstract Syntax Trees (ASTs) and apply rule-based algorithms to detect refactoring and structural edits. Representative tools include RefactoringMiner \cite{Tsantalis:ICSE:2018:RefactoringMiner}, GumTree \cite{DBLP:conf/kbse/FalleriMBMM14}, and ChangeDistiller \cite{Fluri2007ChangeDT}.
RefactoringMiner, for example, introduced a threshold‑free AST‑based matching technique capable of identifying 15 common refactoring types directly from commit histories, without requiring projects to be built. These tools demonstrate the effectiveness of static analysis for structured change detection, but remain constrained by language-specific implementations, fixed taxonomies, and the engineering effort required to adapt them to new ecosystems.

LLMs have demonstrated strong capabilities in code-related tasks such as completion, bug detection, and summarization \cite{brown2020language, kojima2022large, wei2021finetuned}. Code-oriented assistants like Codex \cite{chen2021evaluating} and Claude Code \cite{anthropic_claude_code_2024} leverage large-scale training on code corpora to enable few-shot and zero-shot reasoning. In the context of code review, LLMs have been applied to PR summarization, comment generation, and reviewer assistance \cite{sun2024llms, codedog2024, palos2024feedback}. While these efforts highlight the potential of LLMs to support review workflows, they primarily produce free-text outputs rather than structured representations of code changes.

To the best of our knowledge, no prior work has systematically evaluated LLMs for taxonomy-based labeling of diff hunks in code patches. Existing change-classification tools rely mainly on static analysis, achieving high precision for refactoring detection but remaining language-specific and difficult to customize. In contrast, our work investigates whether LLMs can provide structured, language-agnostic labels for code edits, including cross-hunk relationships (e.g., rename propagation, code moves) and semantic attributes (e.g., old/new identifiers), opening a new direction for automation-friendly code review.

\begin{figure}[t]
  \centering
  \includegraphics[width=0.9\columnwidth]{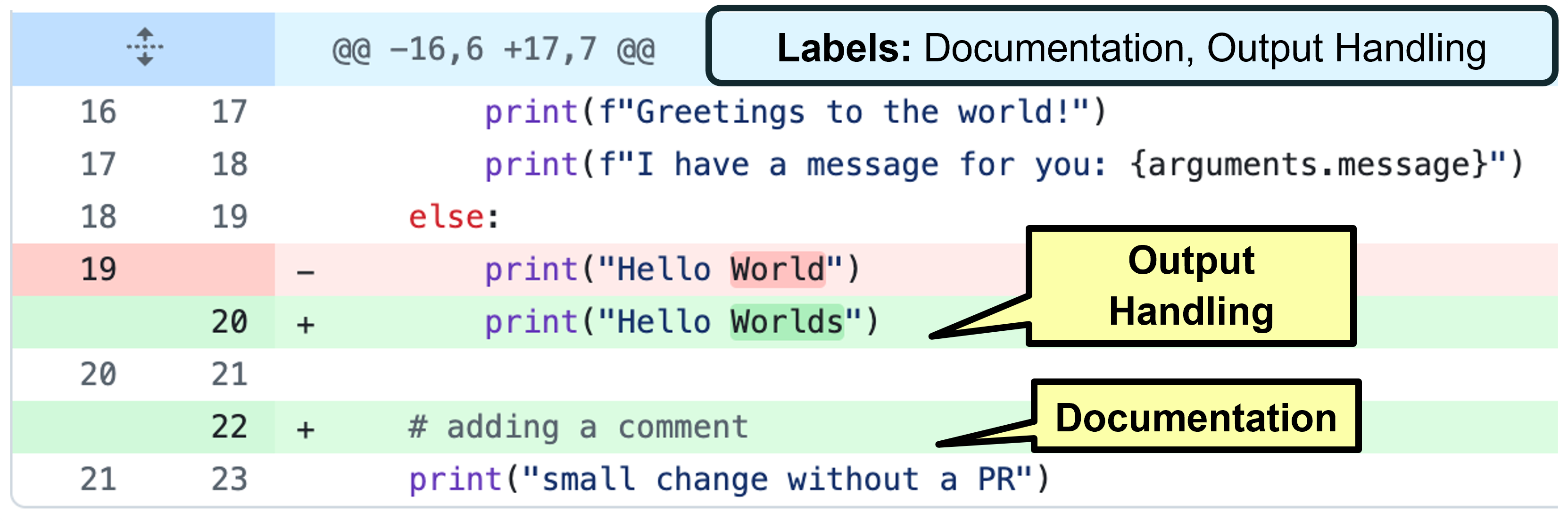}
  \caption{Example of a diff hunk with identified labeling.}
  \label{fig:labeling_illustration}
  \Description{}
\end{figure}

\section{Labeling Code Changes}\label{sec:labeling_patches}
In this paper, we study the problem of labeling code changes. We define a set of label types $\mathcal{T}$, each type describing a supported category of code change (the full taxonomy is detailed in Appendix~\ref{app:label_types}). Using \texttt{git diff}, we decompose a patch into a set of diff hunks $\mathcal{H}$, where each hunk corresponds to a contiguous block of changes between the old and new versions of the code.
Our goal is to assign to each diff hunk in the patch all the label types from $\mathcal{T}$ that describe the changes in it. Formally, a labeling instance is defined as a tuple, $\ell=(i, h, t, p, a)$, where $i \in \mathbb{N}$ is an integer index, $h \in \mathcal{H}$ is the diff hunk the labeling is attached to, $t \in \mathcal{T}$ is the label type, $p \in \mathbb{N}$ is the index of the parent labeling instance, and $a$ is a set of string attributes. For most label types, $p$ and $a$ are unused and are therefore set to default values; their role is described in detail later. The output of our system is the set of all labeling instances, denoted by $\mathcal{L}$. This formulation allows diff hunks to remain unlabeled, indicating that none of the label types in $\mathcal{T}$ adequately describes the changes in the hunk. Importantly, we do not assume that $\mathcal{T}$ exhaustively covers all possible change types. Overall, we wish to assign to each diff hunk \(h \in \mathcal{H}\) a (possibly empty) set of labels, as illustrated in Figure~\ref{fig:labeling_illustration}.

\begin{figure}[t]
  \centering
  \begin{subfigure}{0.64\columnwidth}
        \includegraphics[width=\linewidth]{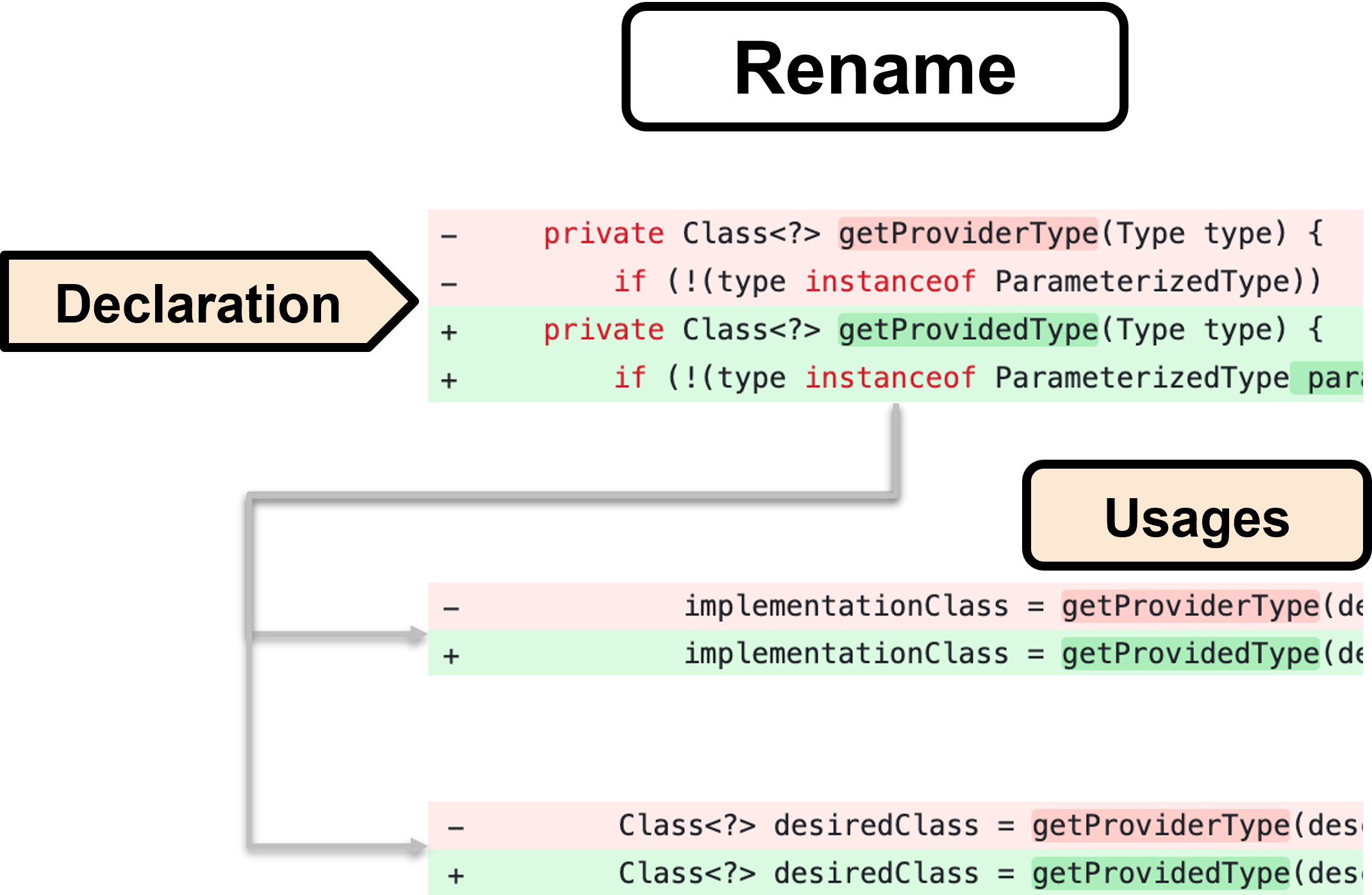}
        \caption{}
        \label{subfig:rename}
    \end{subfigure}
    \hfill
    \begin{subfigure}{0.28\columnwidth}
        \includegraphics[width=\linewidth]{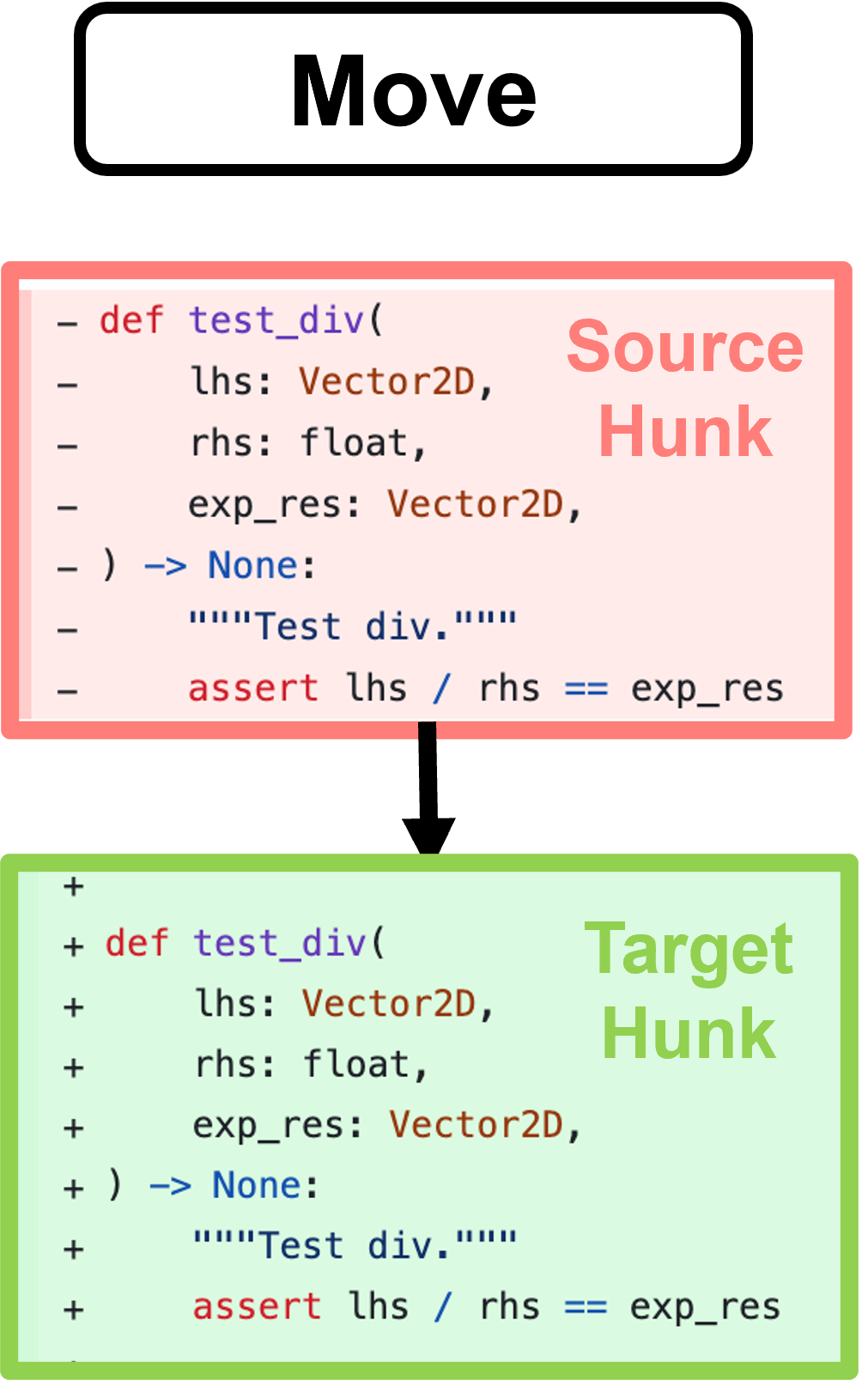}
        \caption{}
        \label{subfig:code_move}
    \end{subfigure}
    \caption{Illustration of hunk relationships in (a) rename and (b) move label types.}
  \label{fig:structure_aware}
  \Description{}
\end{figure}

\begin{figure*}[t]
  \centering
  \includegraphics[width=0.85\textwidth]{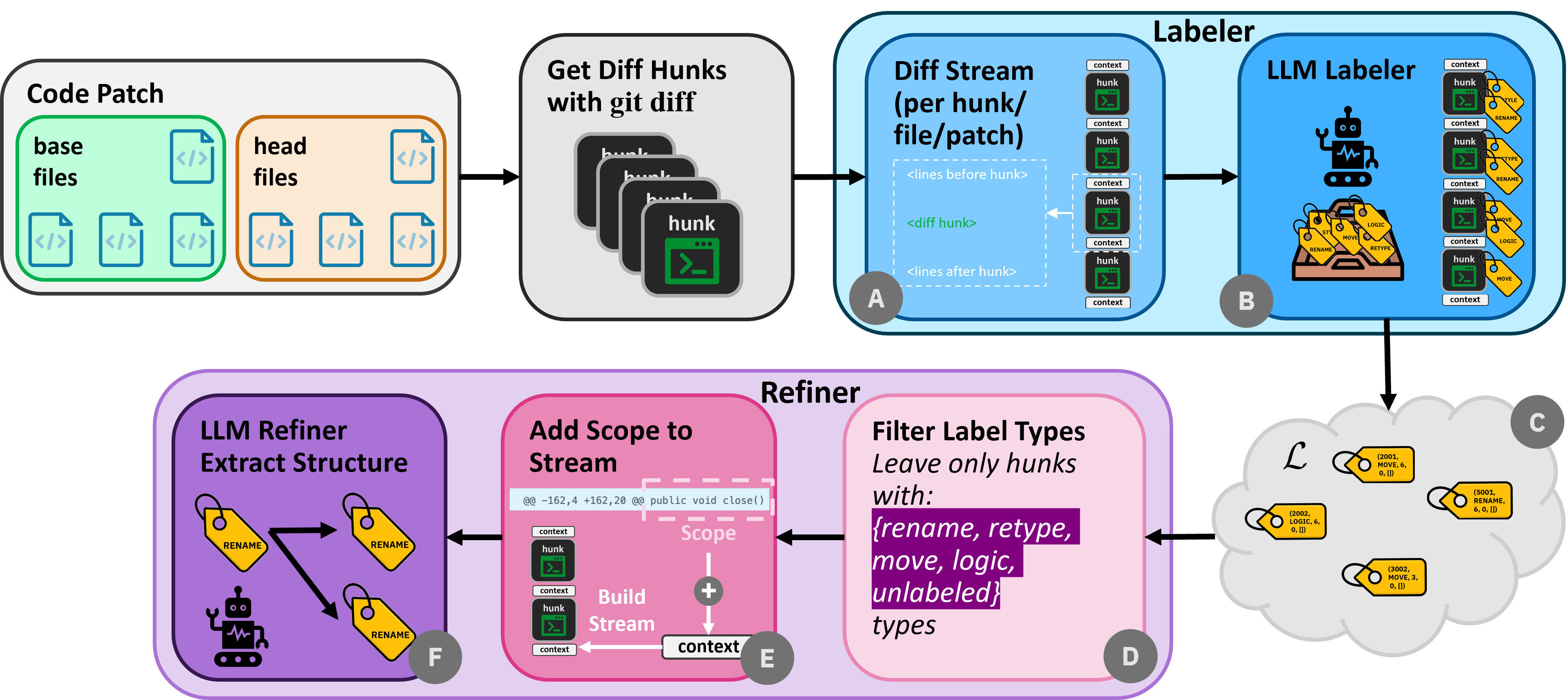}
  \caption{Illustration of our system's pipeline.}
  \label{fig:pipeline}
  \Description{}
\end{figure*}

Static analysis tools can provide information about relationships between edits for certain classes of code changes. We therefore extend our formulation to capture what we term \emph{structure-aware labels}, which explicitly encode relationships between related code changes. Figure~\ref{fig:structure_aware} illustrates this concept. For a \emph{rename} label, the labeling corresponding to the declaration change references the labelings corresponding to the renamed usages. Similarly, for a \emph{move} label, the labeling representing code removal is linked to the labeling representing code addition at the new location. These relationships are encoded via the \(p\) field of the labeling tuple, which stores the index of the parent labeling object, and is set to \(0\) when no such relationship exists. In addition, for \emph{rename} and \emph{retype} labels, we extract attributes describing the change itself. For rename operations, we record the kind of renamed entity (\texttt{VAR}, \texttt{CLASS}, \texttt{PACKAGE}, \texttt{METHOD}, \texttt{ATTRIBUTE}, or \texttt{PARAMETER}), along with the old and new names. For retype operations, we extract the name of the retyped element and its old and new types. These attributes are stored in the \(a\) field, for example \(a = (\texttt{"VAR"}, \texttt{"old\_name"}, \texttt{"new\_name"})\).

\section{Proposed Approach}\label{sec:proposed_approach}
Our objective is to develop an LLM-based tool for labeling patches that can be readily integrated into larger review and analysis pipelines. As outlined in Section \ref{sec:intro}, we adopt a zero-shot and few-shot prompting strategy, following prior work on large language models and in-context learning \cite{brown2020language, wei2021finetuned, kojima2022large}. The pipeline is illustrated in Figure \ref{fig:pipeline}. We will explain its stages throughout this section. Each prompt provided to the LLM includes (i) the label taxonomy definitions, (ii) a task description, (iii) examples for more complex labels, and (iv) a required JSON output schema, together with a stream of diff hunks to be labeled. For each diff hunk, we augment the input with its file name and limited local context—specifically, five non-empty lines of code preceding and following the hunk (Figure \ref{fig:pipeline}A). The model’s JSON response is subsequently parsed into structured output objects for downstream processing.
We decompose the overall problem into two sequential tasks:
(1) Label Assignment, where the LLM assigns zero or more label types to each diff hunk based on its content, and
(2) Structure Extraction, where relationships among labeled hunks are identified.
We refer to the component handling the first task as the \textbf{Labeler}, and to the component performing the second as the \textbf{Label Refiner} (or Refiner for short). 

This decomposition serves several purposes. First, it breaks a complex task into manageable subproblems, improving robustness and simplifying prompt design. Second, it enforces a unified output schema for all labels identified in the Labeler stage. Third, it allows us to vary the context window of the Labeler independently of the Refiner, enabling controlled experimentation with token-budget trade-offs. Lastly, this enables an additional use for the Refiner, correcting errors from the Labeler that can originate from a lack of context in the Labeler.

\subsection{Labeler}\label{subsec:labeler}
As described above, the Labeler assigns label types to individual diff hunks. It operates by prompting an LLM with task instructions and label definitions (see Appendix~\ref{app:labeler_prompt}), followed by a stream of diff hunks. For each hunk \(h \in \mathcal{H}\), the model returns a set of label types \(T(h) \subseteq \mathcal{T}\) (Figure \ref{fig:pipeline}B). These subsets are subsequently converted into the labeling set representation \(\mathcal{L}\) defined in Section~\ref{sec:labeling_patches} (Figure \ref{fig:pipeline}C), with the fields \(p\) and \(a\) initialized to default values and later refined by the Refiner for label types that require that.

Constructing the diff hunk stream introduces a fundamental context trade-off. Providing more hunks supplies the LLM with a richer global context, which can improve label selection; however, longer contexts also increase the burden of filtering irrelevant information, potentially degrading performance. This effect has been observed in both controlled studies~\cite{xu2023retrieval, liu2024lost, hsieh2024ruler} and real-world LLM applications~\cite{yang2024swe}. To study this trade-off, we evaluate three Labeler operating modes: (1) per-hunk inference, (2) per-file inference, and (3) per-patch inference, corresponding to increasing context lengths. Our results empirically demonstrate the impact of this trade-off.

\subsection{Refiner}\label{subsec:refiner}
The Refiner serves two primary purposes: (1) assigning the \(p\) and \(a\) fields to labeling instances, and (2) correcting label types that may have been misassigned by the Labeler due to limited local context. Because labeling instances across files can be related, the Refiner processes the entire patch in a single inference. To bound context length, we restrict its input to diff hunks containing labelings that either require relational fields (rename, retype, move) or may conceal a missed structure-aware label (logic, unlabeled; Figure~\ref{fig:pipeline}D). Accordingly, the Labeler is instructed to conservatively assign the logic label to suspected rename, retype, or move operations when context is insufficient.

Identifying relationships between labeling instances also requires scope awareness. To capture this without substantially increasing context, we include the \texttt{git diff} hunk header before each diff hunk as an approximation of the scope (Figure~\ref{fig:pipeline}E). The Refiner is then prompted with task instructions and label definitions (Appendix~\ref{app:refiner_prompt}) to assign the \(p\) and \(a\) fields (Figure~\ref{fig:pipeline}F). When a single diff hunk contains multiple rename or retype operations, the LLM returns a list of attributes in the \(a\) field, which is subsequently split into distinct labeling instances.

\section{Evaluation}\label{sec:evaluation}
We evaluate our system of a benchmark dataset created in this work. We compare our results on four different LLMs (two open-source and two frontier models): (1) \href{https://swesmith.com/}{SWE-Agent-LM-32B} \cite{yang2025swesmith}, (2) 
\href{https://huggingface.co/meta-llama/Llama-4-Maverick-17B-128E-Instruct}{Llama-4-Maverick-17B-128E-Instruct} \cite{meta_llama4_maverick_17b_128e_2025}, (3) \href{https://www.anthropic.com/news/claude-sonnet-4-5}{Claude Sonnet 4.5} \cite{anthropic_claude_sonnet_4_5_2025}, and (4) \href{https://deepmind.google/models/gemini/pro/}{Gemini-3-Pro-Preview} \cite{google_gemini_3_pro_2025}. We abbreviate these models to SWE, Llama, Sonnet, and Gemini-3, respectively. 
\subsection{Benchmark}\label{subsec:benchmark}
We collected PRs to act as code patches from SWE-bench Multilingual \cite{yang2025swesmith} and SWE-PolyBench \cite{rashid2025swe} and manually labeled them according to the change types present in each diff hunk. To ensure coverage of all label types, we additionally constructed and labeled a set of fabricated PRs. The dataset was split into development and test sets: the development set was used for prompt tuning, while the test set was reserved exclusively for the final evaluation reported in this paper. The test set contains 6 natural and 7 fabricated PRs, comprising 95 diff hunks and 142 labeling instances in total. Most PRs are written in \emph{Java}; however, to reflect the language-agnostic nature of our LLM-based approach (Section~\ref{sec:intro}), we also include \emph{Python} PRs (two in the development set and one in the test set). Additional details about the benchmark are provided in Appendix~\ref{app:bench_stats}.

\subsection{Labeler Results}\label{subsec:labeler_results}
We evaluate the Labeler and Refiner separately in order to isolate their respective contributions. For the Labeler, we compare the predicted label type set $T(h)$ for each diff hunk $h \in \mathcal{H}$ against the manually annotated ground-truth labels $T_{GT}(h)$ from our benchmark. We measure agreement via the intersection of the predicted and ground-truth label sets and define the following two metrics:
\begin{equation}\label{eq:precision_labeler}
    \text{Avg-IoP}(T,T_{GT}) = \frac{1}{|\mathcal{H}|} \sum_{h \in \mathcal{H}} \frac{|
    T(h)\cap T_{GT}(h)|}{|T(h)|} ,
\end{equation}
\begin{equation}\label{eq:recall_labeler}
    \text{Avg-IoGT}(T,T_{GT}) = \frac{1}{|\mathcal{H}|} \sum_{h \in \mathcal{H}} \frac{|T(h)\cap T_{GT}(h)|}{|T_{GT}(h)|} .
\end{equation}
Intersection over Predictor (IoP) measures the fraction of predicted labels that are correct, while Intersection over Ground Truth (IoGT) measures the fraction of ground-truth labels recovered by the Labeler. These correspond to per-hunk precision and recall, respectively; we use the IoP/IoGT terminology to avoid ambiguity with later analyses. Table~\ref{tab:labeler} reports Labeler performance across LLMs and context modes (per-hunk, per-file, and per-patch).

\begin{table}[ht]
\centering
\small
\begin{tabular}{llccc}
\toprule
\textbf{LLM} & \textbf{Mode} & \textbf{Cost [I/O Tokens]} & \textbf{IoP} & \textbf{IoGT} \\
\midrule
\multirow{3}{*}{SWE} & hunk & 1437/77 & 0.63 & 0.62 \\
                         & file & 808/60 & 0.66 & 0.62 \\
                         & patch & 428/51 & 0.7 & 0.63 \\
\midrule
\multirow{3}{*}{Llama} & hunk & 1306/111 & 0.64 & 0.68 \\
                         & file & 803/81 & 0.65 & 0.7 \\
                         & patch & 430/62 & 0.67 & 0.68 \\
\midrule
\multirow{3}{*}{Sonnet} & hunk & 1503/116 & 0.65 & 0.72 \\
                         & file & 946/87 & 0.76 & 0.79 \\
                         & patch & 526/76 & 0.68 & 0.71 \\
\midrule
\multirow{3}{*}{Gemini-3} & hunk & 1436/863 & 0.64 & 0.68 \\
                         & file & 901/373 & \textbf{0.81} & \textbf{0.84} \\
                         & patch & 492/477 & 0.68 & 0.7 \\
\bottomrule
\end{tabular}
\caption{Labeler results for different LLMs and run modes. The cost is presented in input/output tokens used, divided by the number of diff hunks.}
\label{tab:labeler}
\end{table}

\begin{figure}[ht]
  \centering
  \includegraphics[width=0.95\columnwidth]{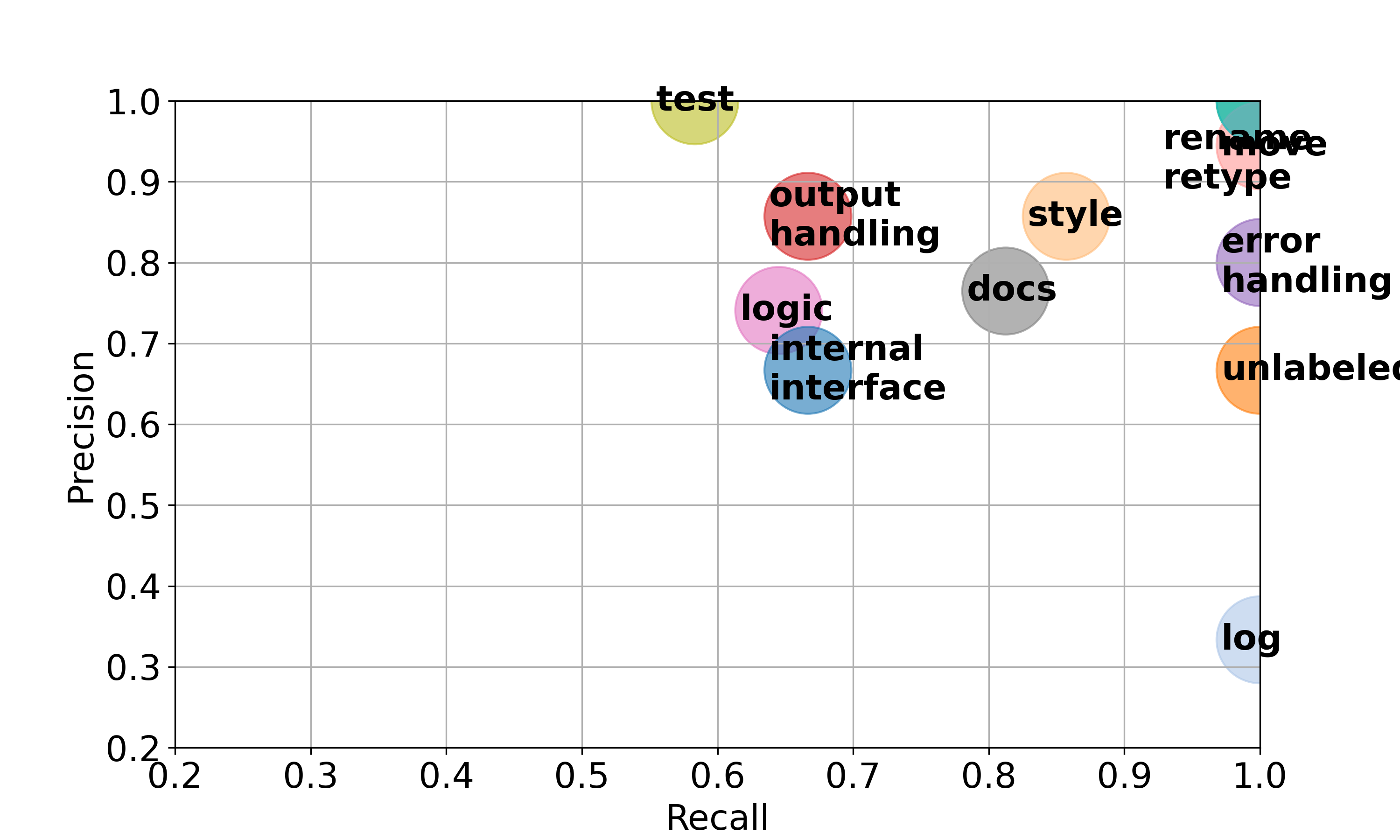}
  \caption{Precision versus recall per label type for the Labeler. Results are shown for Gemini-3 in per-file mode.}
  \label{fig:precision_recall_per_type}
  \Description{}
\end{figure}

We further analyze label difficulty using per-type precision and recall, defined as the fraction of correct predictions and the fraction of recovered ground-truth instances, respectively. Figure~\ref{fig:precision_recall_per_type} shows these metrics for each label type under the best-performing configuration (Gemini-3, per-file mode; Table~\ref{tab:labeler}), with additional results in Appendix~\ref{app:labeler_results}. Performance varies substantially across label types, and the relative ranking of types changes significantly across LLMs. We discuss these trends in Section~\ref{sec:discussion}.
\subsection{Refiner Results}\label{subsec:refiner_results}
The evaluation of the Refiner focuses on the correctness of the attribute field \(a\) and the parent field \(p\) in the labeling objects. We define a \emph{parent-match} between a predicted and a ground-truth labeling object when both reference the same parent diff hunk. For each relevant label type (rename and move), we count—per diff hunk—the number of matching parent references. We then normalize this count by the number of predicted labeling objects of that type to obtain a precision score, and by the number of ground-truth labeling objects to obtain a recall score. We evaluate the attribute field \(a\) in a similar manner for rename and retype labels. Specifically, we count the number of matching attributes between predicted and ground-truth labeling objects within the same diff hunk, and normalize to obtain precision and recall scores. Since each labeling object contains three attributes, we divide by 3 to get a score in the range \([0,1]\). Results for different LLMs and run modes are reported in Table~\ref{tab:refiner}. For the sake of brevity, we present here the results only for the frontier models (Sonnet and Gemini-3) and provide the full results in Appendix~\ref{app:refiner_results}. 

\begin{table}[ht]
\centering
\small
\begin{tabular}{lccc|cc}
\toprule
\textbf{LLM / Mode} & \textbf{Label} & \multicolumn{2}{c|}{\textbf{Attribute}} & \multicolumn{2}{c}{\textbf{Parent}} \\
\cmidrule(lr){3-4} \cmidrule(lr){5-6} 
&  & Precision & Recall & Precision & Recall \\
\midrule
\multirow{3}{*}{Sonnet / hunk} & rename & 0.92 & 0.85 & 0.92 & 0.85 \\
                                   & retype & 0.8 & \textbf{0.67} & - & - \\
                                   & move & - & - & \textbf{1.0} & 0.56 \\
\midrule
\multirow{3}{*}{Sonnet / file} & rename & \textbf{1.0} & 0.85 & \textbf{1.0} & 0.85 \\
                                   & retype & \textbf{1.0} & \textbf{0.67} & - & - \\
                                   & move & - & - & 0.93 & 0.72 \\
\midrule
\multirow{3}{*}{Sonnet / patch} & rename & \textbf{1.0} & 0.85 & \textbf{1.0} & 0.85 \\
                                   & retype & \textbf{1.0} & \textbf{0.67} & - & - \\
                                   & move & - & - & 0.82 & 0.5 \\
\midrule
\multirow{3}{*}{Gemini-3 / hunk} & rename & 0.97 & \textbf{0.9} & \textbf{1.0} & \textbf{0.92} \\
                                   & retype & 0.92 & 0.61 & - & - \\
                                   & move & - & - & \textbf{1.0} & 0.61 \\
\midrule
\multirow{3}{*}{Gemini-3 / file} & rename & 0.72 & 0.67 & \textbf{1.0} & \textbf{0.92} \\
                                   & retype & 0.92 & 0.61 & - & - \\
                                   & move & - & - & 0.93 & \textbf{0.78} \\
\midrule
\multirow{3}{*}{Gemini-3 / patch} & rename & \textbf{1.0} & 0.69 & \textbf{1.0} & 0.69 \\
                                   & retype & \textbf{1.0} & 0.5 & - & - \\
                                   & move & - & - & 0.82 & 0.5 \\
\bottomrule
\end{tabular}
\caption{Refiner results on capturing structure and attributes for different LLMs and run modes, with split precision and recall values}
\label{tab:refiner}
\end{table}

\section{Discussion}\label{sec:discussion}
We now discuss the results presented in Section~\ref{sec:evaluation}. As shown in Table~\ref{tab:labeler}, Gemini-3 achieves the highest overall performance. This improvement, however, comes at a cost: its output token usage is significantly higher than that of the other models, reaching up to 7.5× that of the second-best model, Sonnet. This increase is largely due to the reasoning field in the response schema (Appendix~\ref{app:prompts}), where Gemini-3 provides more detailed explanations, which may partly account for its improved accuracy. For all models except SWE, the per-file inference mode performs best, highlighting a trade-off between context length and labeling accuracy. Consistent with prior work \cite{xu2023retrieval, liu2024lost}, our results suggest that intermediate context sizes offer a better balance between available context and model focus.

The per-type results in Figure~\ref{fig:precision_recall_per_type} indicate that many label types can be detected reliably, though performance varies substantially across LLMs (Appendix~\ref{app:refiner_results}). For most types, precision and recall are correlated; however, for several types, these metrics diverge, creating opportunities for selective downstream use. Label types with high recall but lower precision capture most true instances but also generate false positives, which can be filtered by users or removed through post-processing. Conversely, label types with high precision but lower recall identify predominantly correct instances while missing some true cases; nevertheless, such labels can still be valuable, for example, by highlighting representative changes during codebase exploration or review.

Regarding structure extraction, the results in Table~\ref{tab:refiner} show that relational structure can indeed be captured by LLMs. In particular, Gemini-3 operating in the per-file mode identifies the parent field with very high precision and slightly lower recall. For the attribute field, precision is likewise very high, indicating that when the model identifies a rename or retype instance, its attributes are typically extracted  correctly. Figure~\ref{fig:precision_recall_per_type} further shows that, at the Labeler stage, per-hunk rename and retype detection achieves perfect precision and recall. In contrast, attribute recall remains lower than precision at the Refiner stage. This gap suggests that most errors stem not from incorrect attribute extraction, but from failures to correctly split or enumerate multiple labeling instances within a diff hunk, resulting in fewer extracted instances than required.

Overall, the results demonstrate strong performance, particularly for frontier models. While the approach is not error-free, some degree of imperfection is acceptable in the context of code review. When higher accuracy is required for specific change types, a hybrid strategy can be adopted—using LLM-based labeling for most types while relying on static analysis for a small set of critical label types.

\section{Conclusions and Future Work}\label{sec:conclusions}
We presented a novel LLM-based framework for labeling diff hunks in pull requests. By producing structured outputs, our approach enables systematic evaluation of LLM-based code review methods and supports seamless integration into automated review and analysis pipelines. We further showed that simple cross-hunk relationships, such as renames and code moves, can be captured using our two-stage labeling and refinement pipeline. 

This work also opens several directions for future research. First, the current label taxonomy \(\mathcal{T}\) can be extended beyond primarily syntactic change types to incorporate semantic and intent-level categories, such as performance optimizations, bug fixes, or vulnerability mitigations. Evaluating the reliability of LLMs in identifying such changes and determining the pipeline adaptations required to support them remains an open question. Second, our reliance on \texttt{git diff} hunks may limit granularity, as individual hunks often contain multiple distinct changes. Automatically decomposing large hunks into smaller, semantically coherent units could improve interpretability and labeling accuracy. Finally, we see substantial potential in integrating our structured outputs into hybrid frameworks that combine LLM reasoning with classical static analysis, enabling more robust automated code review systems. We view this work as an initial step toward structured, machine-interpretable LLM-based code review beyond free-text feedback.


\bibliographystyle{ACM-Reference-Format}
\bibliography{bibliography}


\appendix
\clearpage
\section{Supported Code Change Types}\label{app:label_types}
Here we describe the label types set $\mathcal{T}$ supported by our system and their description as it appears in the prompt we provide to the LLM.
\begin{enumerate}
    \item Documentation - adding new or changing existing comments or descriptions. Also include explicit edits of \verb|.txt|, \verb|.md| or similar files.
    \item Testing - changes to testing code.
    \item Output Handling - changes to code that handles \verb|stdout|, \verb|stderr|, writes to output files, print statements, etc.
    \item Retype - changing the type of a variable or attribute. examples: changed \verb|int| to \verb|bool|, as a consequence conditions are different, changed \verb|int| to \verb|long|, return type of a method returns base class rather than inherited class.
    \item Code move - moving code from one location to another, this label should be added at the diff hunk where the code was removed and where it was added. examples: replacing a chunk of code with a function that runs the same code. Moving code from one file to another.
    \item Style change - changes that modify the appearance of the code or the writing style but not the abstract syntax tree (AST). examples: move \verb|{| from same line to line below, change comment style from \verb|//| to \verb|/* */|, split long lines, aligning and indentation (when the indentation does not matter), and other cosmetic changes.
    \item Logging - everything related to logging, initializing the logger, summarizing the log, writing to the log, etc.
    \item Rename - only changes to the name of a variable, method, attribute, class, parameter or package.
    \item Error handling - changes that affect when an error or warning is raised or what happens when they are raised. examples: changes in the try-catch block logic, changes in exception types.
    \item Logic change - any change that modifies the application execution, for example modifies the control flow or results in different application behavior. If you suspect that a diff hunk might be renaming, retyping, or \verb|code_move| but you lack context to decide, label it as \verb|logic_change|.
    \item Internal interface change - The interface of a class or a package are all the publicly accessible elements. Interface changes are changes to the declarations of said elements. The word internal refers to elements that are internal to the application but not internal to a certain file or class. examples: changing methods between being Public or Private, modifying public method declarations or public attributes.
    \item External interface change - changes to the interface itself or user interfaces, the program's external API, command line interface, etc. examples: adding or modifying CLI arguments.
\end{enumerate}

\section{Prompts}\label{app:prompts}
This appendix describes the prompt used for the Labeler and Refiner for the different context modes (per-hunk, per-file, per-patch).
\subsection{Labeler Prompt}\label{app:labeler_prompt}
We describe the structure of the Labeler's Prompt for each of the context modes. We put in purple brackets value placeholders that are filled for the prompts for better readability. The values are given in Appendix \ref{app:placeholders}. Start with the per-hunk prompt structure:




\begin{figure}[h]
\centering
\begin{promptbox}{Labeler per-hunk prompt}

You are an experienced programmer reviewing pull requests on a large GitHub repository.
You are given a diff hunk from a single pull request.
Choose the labels that best describe the changes made in the hunk from the following list:

\textcolor{violet}{\textbf{\{label\_types\}}}

\textcolor{violet}{\textbf{\{specific\_instructions\}}}

Here are some examples to clarify the instructions:

\textcolor{violet}{\textbf{\{examples\}}}

\textcolor{violet}{\textbf{\{hunk\_format\_instructions\}}}

\textcolor{violet}{\textbf{\{json\_format\_request\}}}

Here is the diff hunk and some context:

\textcolor{violet}{\textbf{\{input\_stream\}}}

\end{promptbox}
\label{fig:per_hunk_prompt}
\Description{}
\end{figure}

Now we describe the per-file and per-patch prompt structure. Note that they are the same and will differ only in the length of the input stream.
\begin{figure}[h]
\centering
\begin{promptbox}{Labeler per-patch/file prompt}

You are an experienced programmer reviewing pull requests on a large GitHub repository.
You are given a stream of diff hunks from a single pull request.
For each diff hunk, choose the labels that best describe the changes made in the hunk from the following list:

\textcolor{violet}{\textbf{\{label\_types\}}}

\textcolor{violet}{\textbf{\{specific\_instructions\}}}

---------- Examples (do not respond to these diff hunks) ----------

\textcolor{violet}{\textbf{\{examples\}}}

-------- End of Examples ----------

\textcolor{violet}{\textbf{\{stream\_format\_instructions\}}}

\textcolor{violet}{\textbf{\{json\_format\_request\}}}

Here is the diff hunk stream with some context (respond to all of these):

------------ Diff Hunk Stream -------------

\textcolor{violet}{\textbf{\{input\_stream\}}}

------------ end of Diff Hunk Stream ---------------

\end{promptbox}
\label{fig:per_patch_prompt}
\Description{}
\end{figure}
\subsection{Refiner Prompt}\label{app:refiner_prompt}
Here, we provide the strcture of the Refiner's prompt. For it as well we have placeholder values provided in Appendix \ref{app:placeholders}.
\begin{figure}[h]
\centering
\begin{promptbox}{Refiner prompt}

You are an experienced programmer reviewing pull requests on a large GitHub repository.
You are given a stream of diff hunks from a single pull request that were labeled by another reviewer.
Your task is to provide additional details on the labels.
Here is the description of the labels.

\textcolor{violet}{\textbf{\{label\_types\}}}

For the following labels you should provide attributes and a parent field with the following meaning:

\textcolor{violet}{\textbf{\{parent\_and\_attributes\_instructions\}}}

Note that the parent label might appear after its children in the stream.

The logic change label might be too broad, you may decide to replace it with a more specific label 
using the \verb|updated_type| field in the response and provide attributes and \verb|parent_id| accordingly. 

The \verb|parent_id| must have the same label type as the one pointing to it. 

If you decide to keep the logic change type return the same type provided in the response. 

You should refer to names of attributes of a class using \verb|"<class_name>.<attribute_name>"|.
And to names of arguments of a method using \verb|"<method_name>.<argument_name>"|.
The return value of a method \verb|"<method_name>.return"|.

\textcolor{violet}{\textbf{\{refiner\_stream\_format\_instructions\}}}

\textcolor{violet}{\textbf{\{json\_format\_request\}}}

Here is the diff hunk stream with some context (respond to all of the labels):\\
------------ Diff Hunk Stream -------------\\
\textcolor{violet}{\textbf{\{input\_stream\}}}\\
------------ end of Diff Hunk Stream ---------------
\end{promptbox}
\label{fig:refiner_prompt}
\Description{}
\end{figure}

\subsection{Placeholder values}\label{app:placeholders}
Here we present in table, all the placeholder values from the prompts described in Appendices \ref{app:labeler_prompt} and \ref{app:refiner_prompt}. The \textbf{\{stream\_format\_instructions\}} and \textbf{\{refiner\_stream\_format\_instructions\}} placeholders are too long to fit in a table and hence we provide them separately. The rest of the values are provided in Table \ref{tab:labeler_prompt_placeholders} for the Labeler and Table \ref{tab:refiner_prompt_placeholders} for Refiner specific placeholders.

\paragraph{\textbf{\{stream\_format\_instructions\}}} Ensure the output is valid JSON as it will be parsed using \verb|`json.loads()`| in Python. It should be in the schema: 
\begin{verbatim}
<json>
{
    "response_dict": {
        "<diff_hunk_idx1>": {
        "reasoning": "<reasoning1>",
        "label_names": "[<label1.1>, <label1.2>, ...]"
        },
        "<diff_hunk_idx2>": {
        "reasoning": "<reasoning2>",
        "label_names": "[<label2.1>, <label2.2>, ...]"
        },
        "<diff_hunk_idx3>": {
        "reasoning": "<reasoning3>",
        "label_names": "[<label3.1>, <label3.2>, ...]"
        },
        "<diff_hunk_idx4>": {
        "reasoning": "<reasoning4>",
        "label_names": "[<label4.1>, <label4.2>, ...]"
        }
    }
}
</json>
\end{verbatim}
Where \verb|<diff_hunk_idx1>| is the integer index of the diff hunk, as it is in the diff hunk stream.
And there should be an entry in the \verb|response_dict| for each diff hunk in the stream.
The label names should be exactly the same as in the list above.
Example response:
\begin{verbatim}
<json>
{
    "response_dict": {
        "3": {
            "reasoning": "The diff hunk describes a change in
                the documentation and declares a new public 
            method.",
            "label_names": [
                "internal_interface_change",
                "documentation"
            ]
        },
        "4": {
            "reasoning": "The method my_func was renamed to 
                your_func.",
            "label_names": ["renaming"]
        },
        "5": {
            "reasoning": "The change in the diff hunk
                does not match any of the defined labels.",
            "label_names": []
        }
    }
}
</json>
\end{verbatim}
\begin{table*}[ht]
\centering
\small
\begin{tabular}{|l|p{1.6\columnwidth}|}
\hline
\textbf{Placeholder} & \textbf{Prompt Segment} \\
\hline
\texttt{\{label\_types\}} &
List of label types and descriptions as they are presented in Appendix \ref{app:label_types} in the following format:

label\_name: \verb|{label}|, description: \verb|{description}| \\
\hline
\texttt{\{specific\_instructions\}} &
You may choose more than one label if necessary, a diff hunk might contain more than one type of change.
Compare the diff hunk against all labels in the labels list, and select all the labels 
that fit, considering all the types of changes in the diff hunk.
The labels do not cover all possible code changes.
Therefore, if none of the label describe the change in the diff hunk  
you should decide to assign no labels, i.e., return an empty list of labels.
Pay attention to the file name and its suffix.
 \\
\hline
\texttt{\{examples\}} & \begin{minipage}[t]{\linewidth}
In file \texttt{code.py}:

Example diff hunk:
\begin{verbatim}
 import pandas as pd
 import matplotlib.pyplot as plt
+ import numpy as np
+ import json
+ import os
\end{verbatim}

Example response:
\begin{verbatim}
"reasoning": "The diff hunk describes a change that does
not fit any of the label types in the list, therefore
the response is none of the labels, an empty list."
"label_names": []

\end{verbatim}
\end{minipage}
\\
\hline
\texttt{\{hunk\_format\_instructions\}} &
\begin{minipage}[t]
{\linewidth}
Ensure the output is valid JSON as it will be parsed 
using \verb|`json.loads()`| in Python. 
It should be in the schema: 
\begin{verbatim}
<json>
{
    "reasoning": "<reasoning1>",
    "label_names": "[<label1.1>, <label1.2>, ...]"
}
</json>
\end{verbatim}
The label names should be exactly the same as in the list above. Example response:
\begin{verbatim}
<json>
{
    "reasoning": "The diff hunk describes a change in the
        documentation and declares a new public method.",
    "label_names": ["internal_interface_change", "documentation"]
}
</json>
\end{verbatim}
\end{minipage} \\
\hline
\texttt{\{json\_format\_request\}} &
The output should be only a JSON object according to the schema described above without additional text.
For each diff hunk output every field only once! 
Output only a valid JSON object!
Do not include quotation marks inside the reasoning string as they will make 
the JSON invalid, for example:
Don't say \verb|in the print statement print("hello")|, instead say \verb|print(hello)| or
simply don't quote such statements.
Do not start the JSON with \verb|```json| or end with \verb|```|, 
instead use \verb|<json>| at the beginning and \verb|</json>| at the end of the JSON. \\
\hline
\end{tabular}
\caption{Placeholder fields used in the Labeler's prompt templates.}
\label{tab:labeler_prompt_placeholders}
\end{table*}

\paragraph{\textbf{\{refiner\_stream\_format\_instructions\}}} Ensure the output is valid JSON as it will be parsed 
using `json.loads()` in Python. 
It should be in the schema: 
\begin{verbatim}
<json>
{
    "response_dict": {
        "<label_id1>": {
        "reasoning": "<reasoning1>",
        "updated_type": <updated_type1>,
        "attributes": [<attribute1.1>, 
            <attribute1.2>, <attribute1.3>],
        "parent_id": <parent_id1>
        },
        "<label_id2>": {
        "reasoning": "<reasoning2>",
        "updated_type": <updated_type2>,
        "attributes": [<attribute2.1>, 
            <attribute2.2>, <attribute2.3>],
        "parent_id": <parent_id2>
        },
        "<label_id3>": {
        "reasoning": "<reasoning3>",
        "updated_type": <updated_type3>,
        "attributes": [],
        "parent_id": <parent_id3>
        },
        "<label_id4>": {
        "reasoning": "<reasoning4>",
        "updated_type": <updated_type4>,
        "attributes": [<attribute4.1>,
            <attribute4.2>, <attribute4.3>],
        "parent_id": <parent_id4>
        },
    }
}
</json>
\end{verbatim}
Where <label\_id> is the integer index of the label, as it is in the input stream.
And there should be an entry in the response\_dict for each label in the stream.
The label names should be exactly the same as in the list above.
Example response:
\begin{verbatim}
<json>
{
    "response_dict": {
        "1002": {
        ""reasoning": "The orignal label fits this 
        change becuase...",
        "updated_type": "LOGIC_CHANGE",
        "attributes": [],
        "parent_id": "0"
        },
        "3000": {
        "reasoning": "This label is a consequence 
        of the RENAME label with id 5002.",
        "updated_type": "RENAME",
        "attributes": ["METHOD", "my_func", 
            "your_func"],
        "parent_id": "5002"
        },
        "5001": {
        "reasoning": "Code from this diff hunk was 
            moved to diff hunk 8. as mentioned by 
            label 8000",
        "updated_type": "CODE_MOVE",
        "attributes": [],
        "parent_id": "8000"
        },
        "5002": {
        "reasoning": "The name of the method 
            my_func was renamed to your_func 
            in its declaration. ",
        "updated_type": "RENAME",
        "attributes": ["METHOD", "my_func", 
            "your_func"],
        "parent_id": "0"
        },
        "7001": {
        "reasoning": "This label is a consequence 
            of the RENAME label with id 5002.",
        "updated_type": "RENAME",
        "attributes": ["METHOD", "my_func", 
            "your_func"],
        "parent_id": "5002"
        },
        "9003": {
        "reasoning": "This label is a consequence 
            of the RENAME label with id 5002.",
        "updated_type": "RENAME",
        "attributes": ["VAR", "x", "y"],
        "parent_id": "5002"
        },
        "8000": {
        "reasoning": "Code from diff hunk 5 was 
            moved here as mentioned by label 5001",
        "updated_type": "CODE_MOVE",
        "attributes": [],
        "parent_id": "0"
        },
        9002: {
        "reasoning": "the variable x was changed 
            from type int to long",
        "updated_type": "RETYPE",
        "attributes": ["x", "int", "long"],
        "parent_id": "0"
        }
    }
}
</json>
\end{verbatim}

\paragraph{Input Stream Format}
This is a formatting of the diff hunks processed in this call to the LLM. It changes based on the input but it has the following structure. The Labeler's \textbf{per-hunk} stream follows the format below:
\begin{verbatim}
In file <filename>:
Code above the diff hunk:
```
<5 lines before hunk>
```
Diff hunk content:
Header <header>:
```
<diff hunk>
```
Code below the diff hunk:
<5 lines after diff hunk>
\end{verbatim}
And the \textbf{per-file} input stream has the following format:
\begin{verbatim}
In file <file_name>:
Diff hunk number <diff_hunk_index_1>:
```
<5 lines before hunk 1>
<diff_hunk_1>
<5 lines after hunk 1>
```
.
.
.
Diff hunk number <diff_hunk_index_k>:
```
<5 lines before hunk k>
<diff_hunk_k>
<5 lines after hunk k>
```
\end{verbatim}
The \textbf{per-patch} format simply is a concatenation of the per-file format. The \textbf{Refiner}'s stream follows the format below:
\begin{verbatim}
In file <filename>:
Diff hunk number <diff hunk_index_1> in scope <scope_1>:
Labeled as:
Type: <type_1.1>, ID: <label_id_1.1>
.
.
.
Type: <type_1.m>, ID: <label_id_1.m>
```
<5 lines before hunk 1>
<diff_hunk_1>
<5 lines after hunk 1>
```
.
.
.
\end{verbatim}
And this format continues for all files and diff hunks. 
\begin{table*}[ht]
\centering
\small
\begin{tabular}{|l|p{1.6\columnwidth}|}
\hline
\textbf{Placeholder} & \textbf{Prompt Segment} \\
\hline
\texttt{\{parent\_and\_attributes} & \\ \texttt{\_instructions\}} &
\begin{minipage}[t]
{\linewidth}
\begin{verbatim}
RENAME:
    parent_id - each rename is either a change in the declaration or a change to the usage
    i.e., a consequence of the change in the declaration. 
    If the change is in the declaration this is the root of the change and the parent should be 0. 
    If the change is a consequence the parent should contain the label id of the declaration change.
    
    attributes - for each rename we require three fields with the following order:
    
    The type of element that was renamed, one of the following:
    
    VAR, ATTRIBUTE, METHOD, CLASS, PARAMETER, PACKAGE.
    The original name.
    The new name. 
    
    Example: if there are two renames described by a label the attributes should be:
    ["VAR", "my_var", "your_var", "CLASS", "MyClass", "YourClass"]    
    
RETYPE:
    parent_id - always 0
    attributes - for each retype we require three fields with the following order:
    The name of the retyped element.
    The original type.
    The new type. 
    Example: if there are two retypes described by a label the attributes should be:
        ["x", "float", "complex", "res", "list[str]", "dict[str, str]"]

CODE_MOVE:
    parent_id - Where the code was added the parent is 0 and where it was removed
    it should be the label id of the label that describes the addition of the matching code.
    attributes - None, return an empty list []

LOGIC_CHANGE:
    parent_id - always 0
    attributes - None, return an empty list []  
\end{verbatim}
\end{minipage} \\
\hline
\end{tabular}
\caption{Placeholder fields used in the Refiner's prompt template.}
\label{tab:refiner_prompt_placeholders}
\end{table*}

\begin{figure*}[h]
  \centering
  \begin{subfigure}{0.4\textwidth}
        \includegraphics[width=\linewidth]{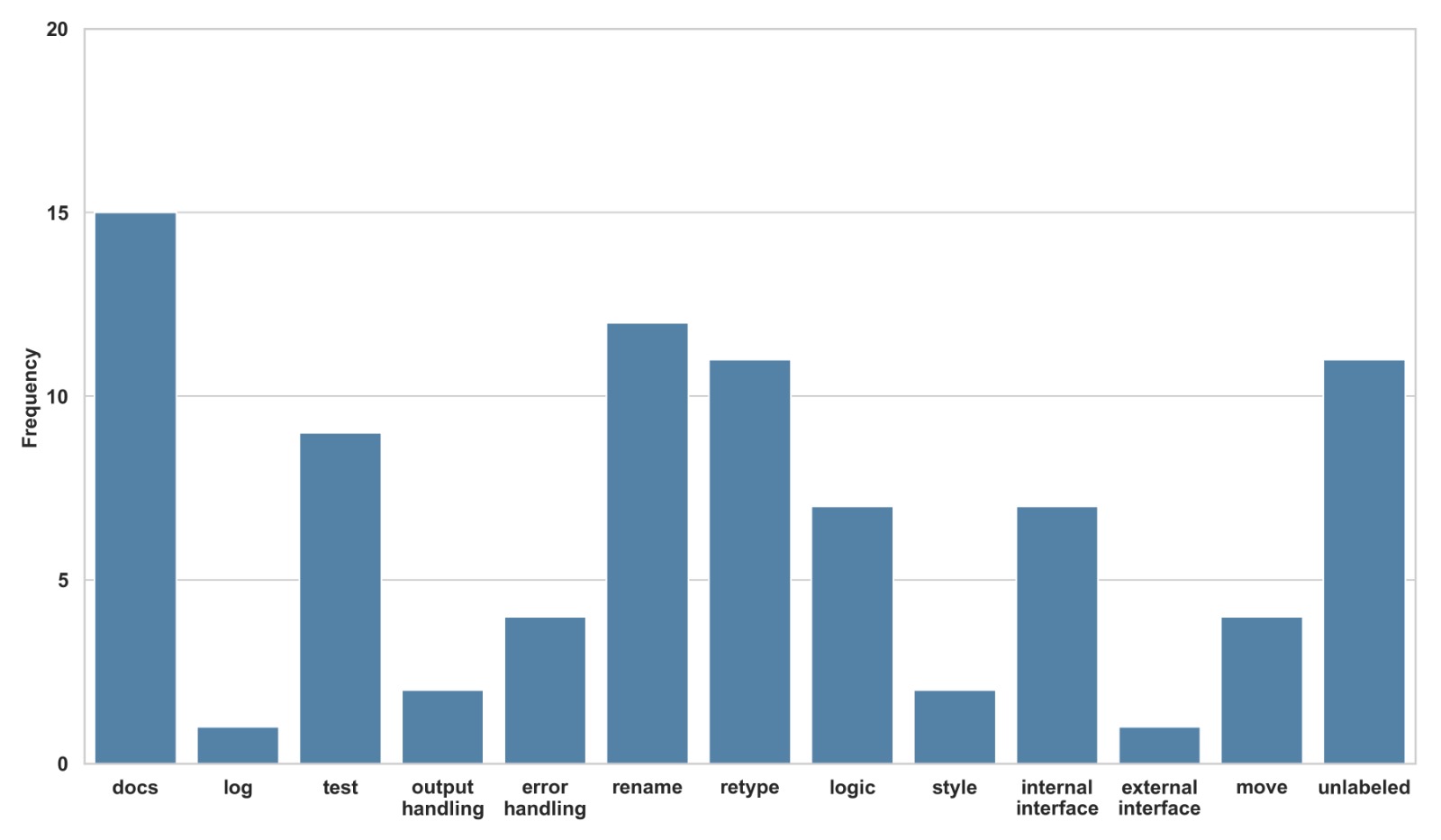}
        \caption{}
        \label{subfig:validation_hist}
    \end{subfigure}
    \hfill
    \begin{subfigure}{0.4\textwidth}
        \includegraphics[width=\linewidth]{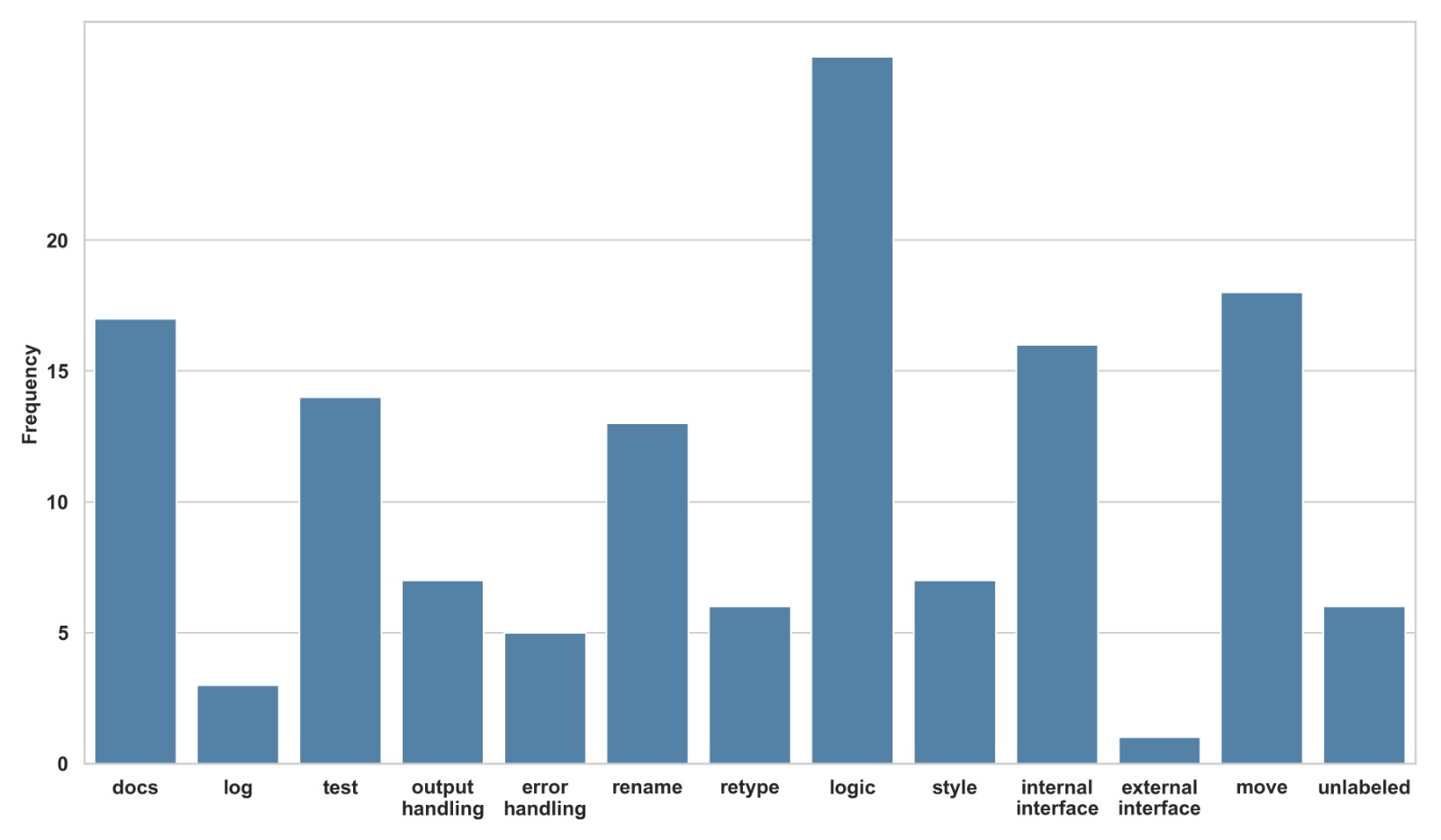}
        \caption{}
        \label{subfig:test_hist}
    \end{subfigure}
    \caption{Histogram of label occurrences in our benchmark's (a) validation set and (b) test set.}
  \label{fig:benchmark_hist}
  \Description{}
\end{figure*}

\section{Evaluation Appendix}
Here, we provide additional details on the evaluation in Section \textcolor{cyan}{5}. 
\subsection{Benchmark Statistics}\label{app:bench_stats}
Here, we present additional details on the benchmark we constructed for this paper. As mentioned in Section \textcolor{cyan}{5.1}, we split the data into development and test sets. The development set contains 3 natural \emph{JAVA} PRs, 2 fabricated \emph{Python} PRs. The histogram of label type appearances in the development set and test set are shown in Figure \ref{fig:benchmark_hist}. Importantly, all label types are represented in both sets and we made an effort to keep the type frequencies as similar as possible. 

\subsection{Additional Labeler Results}\label{app:labeler_results}
Here, we present additional results for the Labeler. We present the IoP and IoGT of the Labeler without the type update by the refiner in Table \ref{tab:labeler_only}. For almost all modes there is a slight improvement becuase of the additional refiner step, but the trend of per-file being the best performing mode remains. And Gemini-3 is still the best performing model.  
\begin{table}[ht]
\centering
\small
\begin{tabular}{llccc}
\toprule
\textbf{LLM} & \textbf{Mode} & \textbf{Cost [I/O Tokens]} & \textbf{IoP} & \textbf{IoGT} \\
\midrule
\multirow{3}{*}{SWE} & hunk & 1437/77 & 0.63 & 0.62 \\
                         & file & 808/60 & 0.64 & 0.62 \\
                         & patch & 428/51 & 0.67 & 0.61 \\
\midrule
\multirow{3}{*}{Llama} & hunk & 1306/111 & 0.58 & 0.63 \\
                         & file & 803/81 & 0.64 & 0.69 \\
                         & patch & 430/62 & 0.67 & 0.69 \\
\midrule
\multirow{3}{*}{Sonnet} & hunk & 1503/116 & 0.62 & 0.67 \\
                         & file & 946/87 & 0.75 & 0.77 \\
                         & patch & 526/76 & 0.68 & 0.71 \\
\midrule
\multirow{3}{*}{Gemini-3} & hunk & 1436/863 & 0.60 & 0.62 \\
                         & file & 901/373 & \textbf{0.79} & \textbf{0.83} \\
                         & patch & 492/477 & 0.69 & 0.71 \\
\bottomrule
\end{tabular}
\caption{Labeler results for different LLMs and run modes. The cost is presented in input/output tokens used, divided by the number of diff hunks.}
\label{tab:labeler_only}
\end{table}

We also present here the per-type precision recall plots for additional run modes in Figure \ref{fig:type_precision_recall_llms}. We show results for Gemini-3 and Sonnet in per-hunk and per-file modes. While some label types appear to have consistent results (rename, retype, docs).

\subsection{Refiner Results}\label{app:refiner_results}
\begin{figure*}[b]
  \centering
  \begin{subfigure}{0.45\textwidth}
        \includegraphics[width=\linewidth]{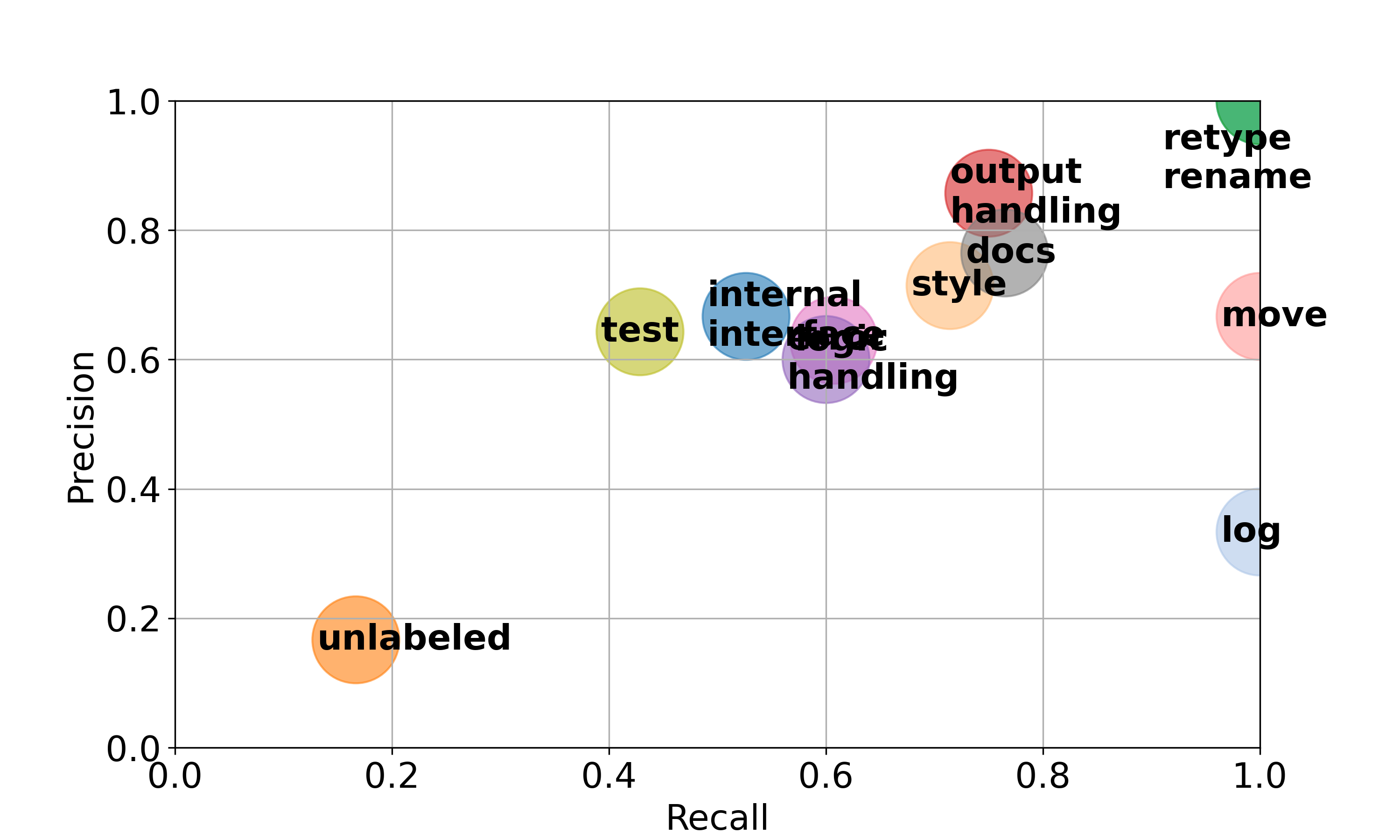}
        \caption{}
        \label{subfig:dh_gemini}
    \end{subfigure}
    \hfill
    \begin{subfigure}{0.45\textwidth}
        \includegraphics[width=\linewidth]{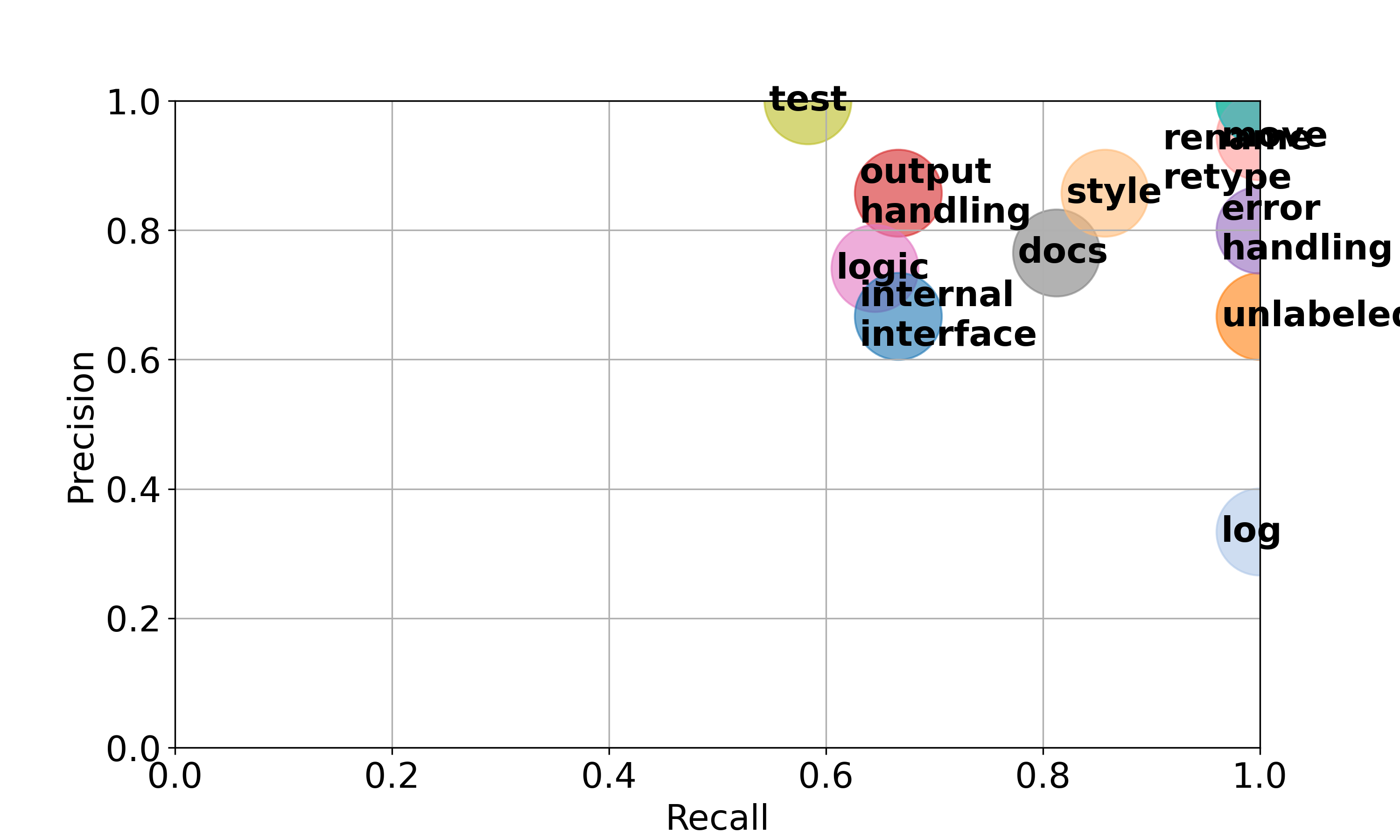}
        \caption{}
        \label{subfig:file_gemini}
    \end{subfigure}
    \begin{subfigure}{0.45\textwidth}
        \includegraphics[width=\linewidth]{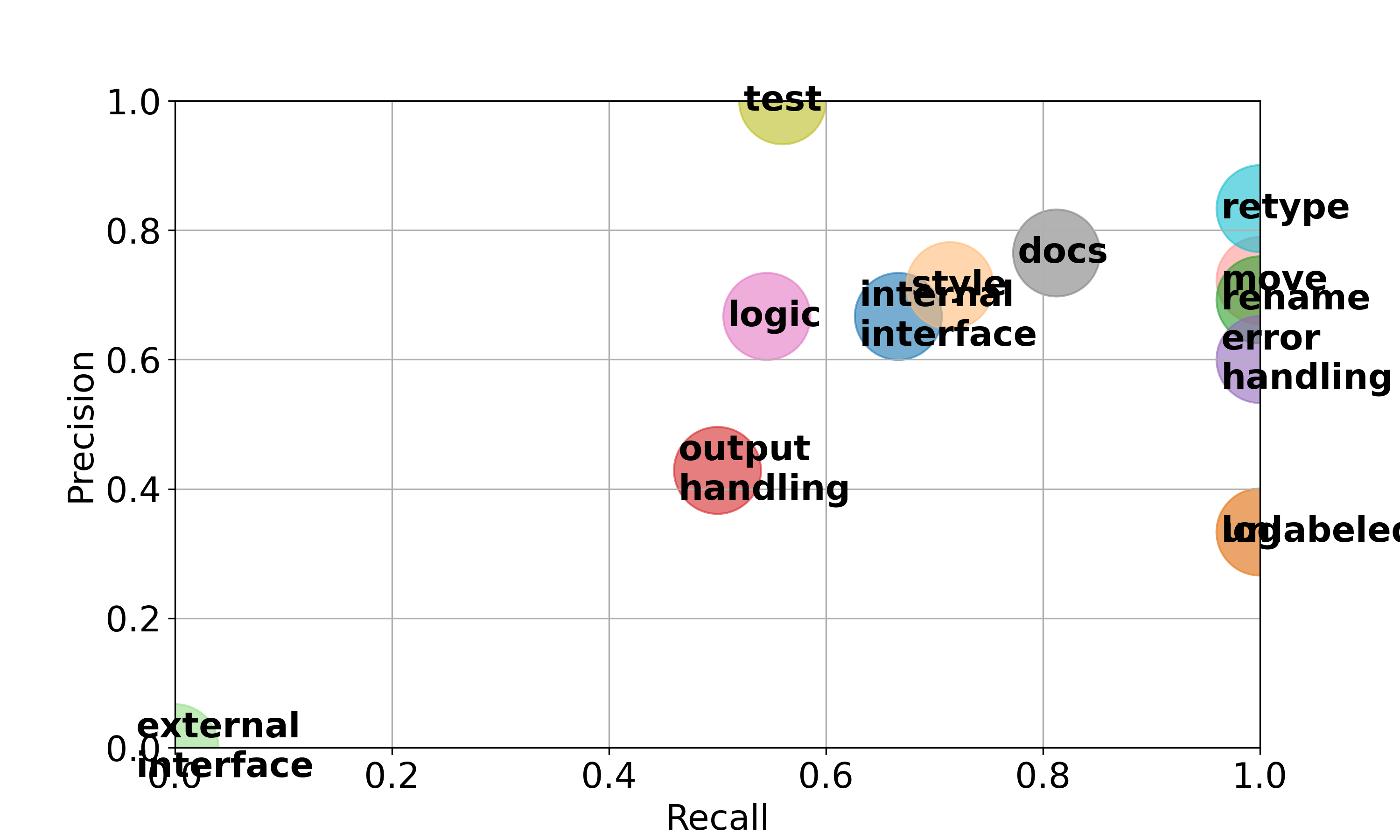}
        \caption{}
        \label{subfig:pr_gemini}
    \end{subfigure}
    \hfill
    \begin{subfigure}{0.45\textwidth}
        \includegraphics[width=\linewidth]{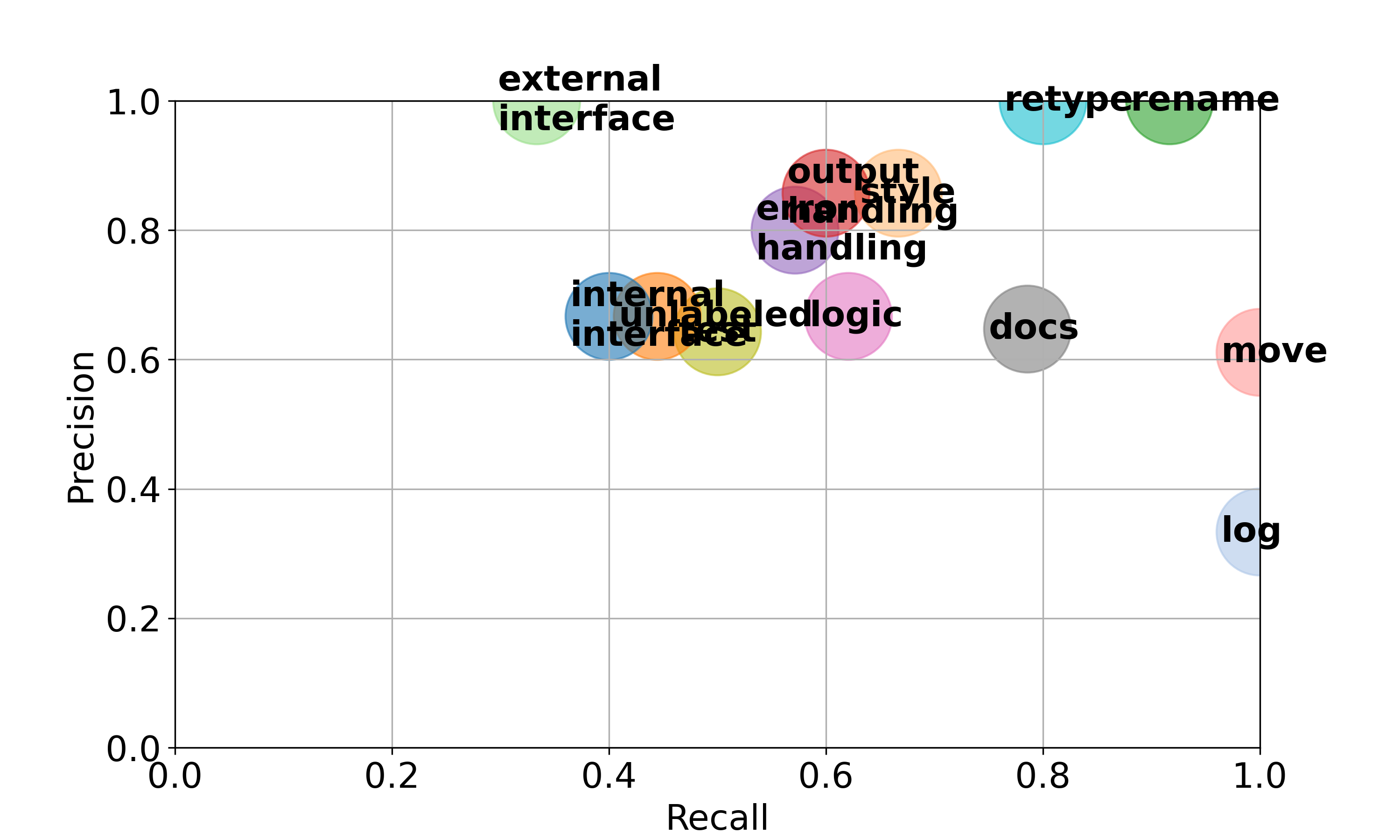}
        \caption{}
        \label{subfig:df_sonnet}
    \end{subfigure}
    \begin{subfigure}{0.45\textwidth}
        \includegraphics[width=\linewidth]{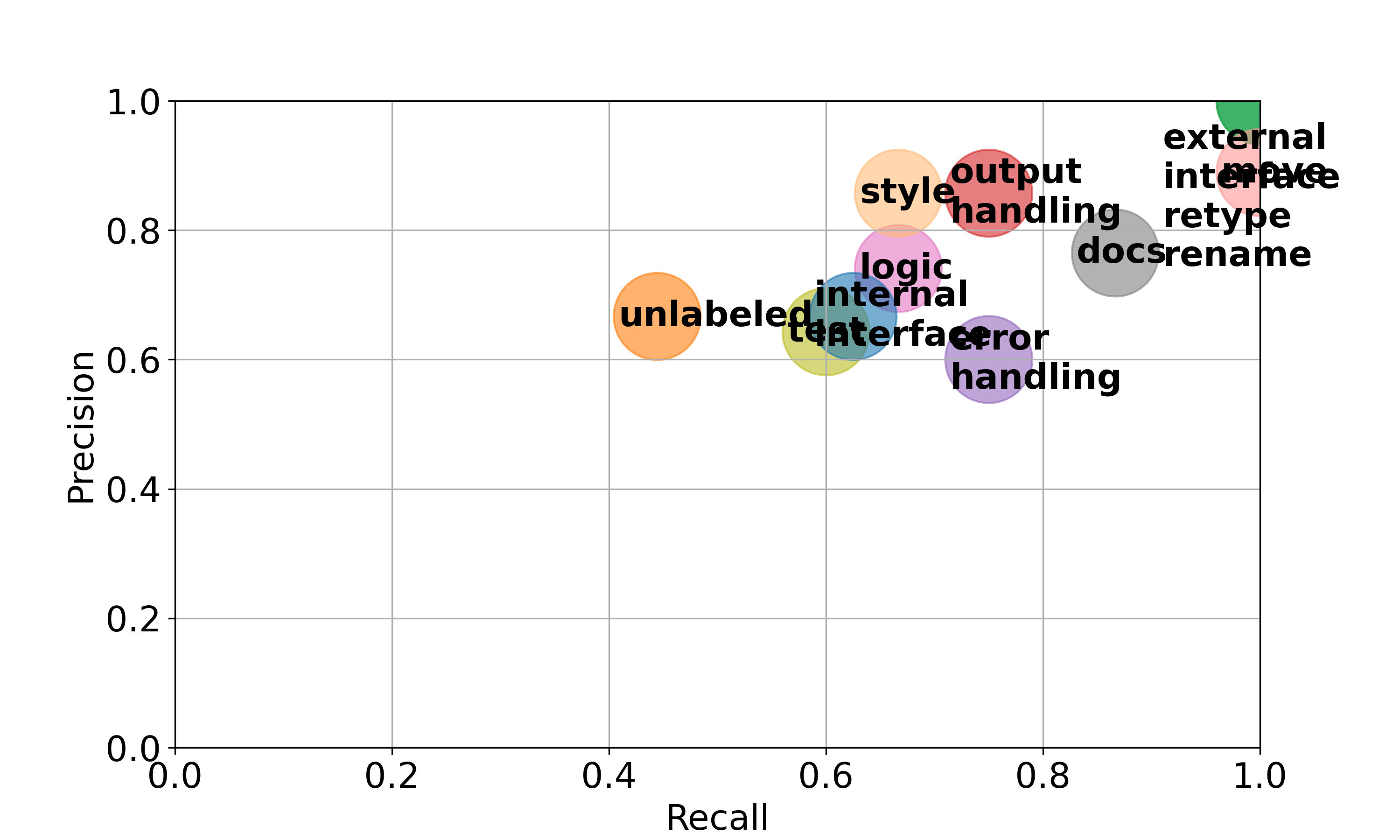}
        \caption{}
        \label{subfig:file_sonnet}
    \end{subfigure}
    \hfill
    \begin{subfigure}{0.45\textwidth}
        \includegraphics[width=\linewidth]{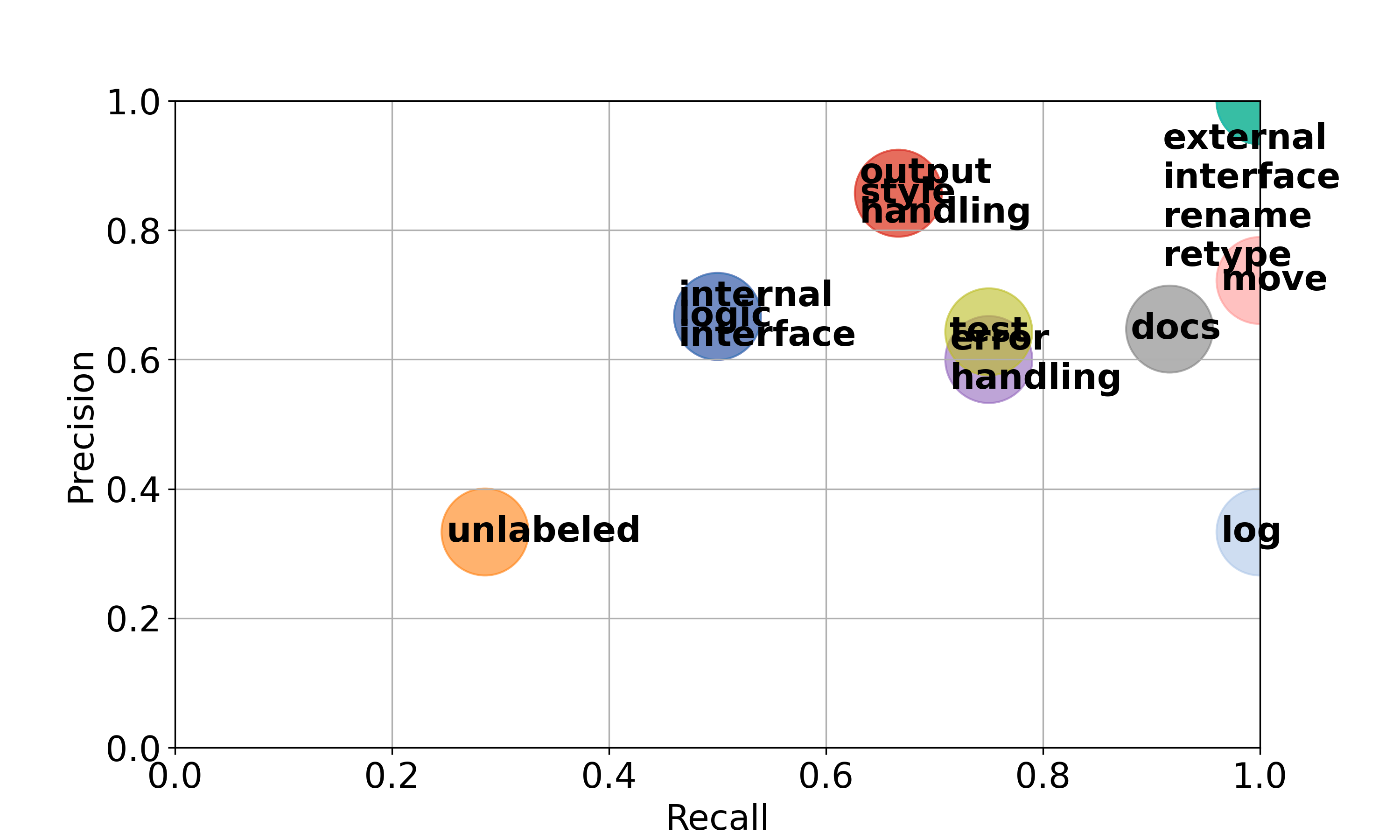}
        \caption{}
        \label{subfig:pr_sonnet}
    \end{subfigure}
    \caption{Per-type precision recall plots for (a) Gemini-3 per-hunk, (b) Gemini-3 per-file, (c) Sonnet per-hunk, and (d) Sonnet per-file.}
  \label{fig:type_precision_recall_llms}
  \Description{}
\end{figure*}

We present here the Refiner results for all models in Table \ref{tab:all_refiner}. We further present the per-type precision vs. recall plots for the frontier models in all inference modes (per-hunk, per-file and per-patch) in Figure \ref{fig:type_precision_recall_llms}. As mentioned in the main text, the performance of each type for the different models and inference modes varies significantly. This is especially true for label types with less instances in the benchmark's test set (external interface, error handling, unlabeled, log). This fact makes it difficult to draw conclusions on the behavior of each type, and a larger scale experiment is required before deciding how the handle errors of the different types.

\begin{table}[ht]
\centering
\small
\begin{tabular}{lccc|cc}
\toprule
\textbf{LLM / Mode} & \textbf{Label} & \multicolumn{2}{c|}{\textbf{Attribute}} & \multicolumn{2}{c}{\textbf{Parent}} \\
\cmidrule(lr){3-4} \cmidrule(lr){5-6} 
&  & Precision & Recall & Precision & Recall \\
\midrule
\multirow{3}{*}{SWE / hunk} & rename & 0.67 & 0.62 & 0.92 & 0.85 \\
                                   & retype & 0.52 & 0.61 & - & - \\
                                   & move & - & - & 0.0 & 0.0 \\
\midrule
\multirow{3}{*}{SWE / file} & rename & 0.52 & 0.52 & 0.56 & 0.56 \\
                                   & retype & 0.75 & 0.6 & - & - \\
                                   & move & - & - & 0.1 & 0.06 \\
\midrule
\multirow{3}{*}{SWE / patch} & rename & 1.67 & 1.54 & 0.58 & 0.54 \\
                                   & retype & 0.8 & \textbf{0.67} & - & - \\
                                   & move & - & - & 0.625 & 0.28 \\
\midrule
\multirow{3}{*}{Llama / hunk} & rename & 0.63 & 0.77 & 0.81 & \textbf{1.0} \\
                                   & retype & 0.67 & 0.44 & - & - \\
                                   & move & - & - & 0.375 & 0.17 \\
\midrule
\multirow{3}{*}{Llama / file} & rename & 0.58 & 0.49 & \textbf{1.0} & 0.85 \\
                                   & retype & 0.67 & 0.56 & - & - \\
                                   & move & - & - & 0.44 & 0.22 \\
\midrule
\multirow{3}{*}{Llama / patch} & rename & 0.63 & 0.44 & 0.33 & 0.23 \\
                                   & retype & 0.43 & 0.5 & - & - \\
                                   & move & - & - & 0.64 & 0.39 \\
\midrule
\multirow{3}{*}{Sonnet / hunk} & rename & 0.92 & 0.85 & 0.92 & 0.85 \\
                                   & retype & 0.8 & \textbf{0.67} & - & - \\
                                   & move & - & - & \textbf{1.0} & 0.56 \\
\midrule
\multirow{3}{*}{Sonnet / file} & rename & \textbf{1.0} & 0.85 & \textbf{1.0} & 0.85 \\
                                   & retype & \textbf{1.0} & \textbf{0.67} & - & - \\
                                   & move & - & - & 0.93 & 0.72 \\
\midrule
\multirow{3}{*}{Sonnet / patch} & rename & \textbf{1.0} & 0.85 & \textbf{1.0} & 0.85 \\
                                   & retype & \textbf{1.0} & \textbf{0.67} & - & - \\
                                   & move & - & - & 0.82 & 0.5 \\
\midrule
\multirow{3}{*}{Gemini-3 / hunk} & rename & 0.97 & \textbf{0.9} & \textbf{1.0} & 0.92 \\
                                   & retype & 0.92 & 0.61 & - & - \\
                                   & move & - & - & \textbf{1.0} & 0.61 \\
\midrule
\multirow{3}{*}{Gemini-3 / file} & rename & 0.72 & 0.67 & \textbf{1.0} & 0.92 \\
                                   & retype & 0.92 & 0.61 & - & - \\
                                   & move & - & - & 0.93 & \textbf{0.78} \\
\midrule
\multirow{3}{*}{Gemini-3 / patch} & rename & \textbf{1.0} & 0.69 & \textbf{1.0} & 0.69 \\
                                   & retype & \textbf{1.0} & 0.5 & - & - \\
                                   & move & - & - & 0.82 & 0.5 \\
\bottomrule
\end{tabular}
\caption{Refiner results on capturing structure and attributes for different LLMs and run modes, with split precision and recall values}
\label{tab:all_refiner}
\end{table}

\end{document}